\documentclass[12pt,a4paper]{article}
\usepackage{amssymb,amsfonts,amsmath}

\title{CHSH inequality: Quantum probabilities as classical conditional probabilities}

\author{Andrei Khrennikov\\
International Center for Mathematical Modeling\\
 in Physics and Cognitive Sciences \\
 Linnaeus University,  V\"axj\"o-Kalmar, Sweden}

\date{}
 
\begin{document}

\maketitle

\abstract{The celebrating theorem of A. Fine implies that the CHSH inequality is violated if and only if the joint probability distribution 
for the quadruples of observables involved the EPR-Bohm-Bell experiment does not exist, i.e., it is impossible to use the classical probabilistic 
model (Kolmogorov, 1933). In this note we demonstrate that, in spite of Fine's theorem, the results of observations in the EPR-Bohm-Bell experiment 
can be described in the classical probabilistic framework. However, the ``quantum probabilities'' have to be interpreted as conditional 
probabilities, where conditioning is with respect to fixed experimental
settings. Our approach is based on the complete account of randomness involved in the experiment. The crucial point is 
that randomness of selections of experimental settings has to be taken into account. This approach can be applied to any complex experiment
in which statistical data are collected for various (in general incompatible) experimental settings. Finally, we emphasize that 
our construction of the classical probability
space for the EPR-Bohm-Bell experiment cannot be used to support the hidden variable approach to the quantum phenomena. The classical 
random parameter $\omega$ involved in our considerations cannot be identified with the hidden variable $\lambda$ which is used 
the Bell-type considerations.}

\section{Introduction}

Although this year we celebrate the 50th anniversary  of Bell's inequality \cite{Bell0} (see also \cite{Bell}), its interpretations and, in particular, interpretations
of its probabilistic structure are still the hot topic of discussions on foundations of quantum mechanics and quantum information 
theory, e.g., in \cite{2}--\cite{6}. Since Bell's types inequalities play the fundamental role in various applications of quantum information, especially in quantum 
cryptography and theory of quantum random generators, the debates on the Bell inequality have important consequences for justification 
of modern quantum technologies (see again, e.g.,  \cite{2}--\cite{6}). We also remark that recently essential progress was approached 
in the performance of a loophole free Bell test \cite{BT1}, \cite{BT2} (see also \cite{BT3}),
although it is still unclear when such a final test will be finally performed. 

There is the common opinion that violation of the Bell-type inequalities by quantum correlations  
implies that the laws of classical probability theory (based on 
the Kolmogorov measure-theoretic axiomatics \cite{K}, 1933) cannot be applied to the 
description of quantum phenomena (at least for entangled systems).  The heuristic roots of such 
a viewpoint are clear, cf. \cite{A1}--\cite{18}. The statistical data used to violate the Bell-type inequalities are collected for 
pairs of incompatible experimental settings. Since in  the Kolmogorov model all observables have to be
represented by random variables (measurable functions) on the same sample space, it is reasonable to expect
that such a construction can´not be used in the case of incompatibility, see, e.g., 
Such heuristic reasoning can be mathematically justified with aid the  celebrating theorem of A. Fine \cite{Fine}.
This theorem states that the CHSH-inequality \cite{CHSH} is violated if and only if {\it the joint probability distribution 
for the quadruples of observables involved the EPR-Bohm-Bell experiment does not exist}, 
i.e., it is impossible to use the classical probabilistic 
model (Kolmogorov, 1933).

In this note we demonstrate that, in spite of the Fine theorem the results of observations in the EPR-Bohm-Bell experiment 
can be described in the classical probabilistic framework. However, the {\it ``quantum probabilities'' have to be interpreted as conditional 
probabilities}, where conditioning is with respect to fixed experimental
settings.  Our approach is based on the complete account of randomness involved in the experiment. The crucial point is 
that randomness of selections of experimental settings has to be taken into account. 
In this paper we present the general Kolmogorovian construction for complex experiments combing a few (in general incompatible) 
experimental settings; the concrete model of the Kolmogorov space was presented in \cite{Av1}, \cite{Av2}.  We emphasize 
that our construction can be used for data of any kind, i.e., not only for data from quantum experiments \cite{UB}, cf. also 
with the recent papers of Dzhafarov and Kujala \cite{DZQ}, \cite{DZQ1}. We point out that the problem of embedding of quantum statistical data into the classical probabilistic 
model has interesting couplings to the problem of freewill \cite{Av2}, cf. with the paper of Kofler  et al. \cite{Kofler2}.

We remark that the problem of inter-relation between quantum and classical probability was studied in hundreds of papers. 
Here we even do not try to present the corresponding bibliography, see, e.g., \cite{ENTROPY}, \cite{book2} for reviews; 
besides of cited papers related to the Bell inequality, we can 
mention the works of  De Gosson, e.g.,  \cite{Gosson},  \cite{Gosson1} 
on inter-relation between probabilistic structures of classical and quantum mechanics
and a series of papers of  Manko et al., see, e.g.,  \cite{M0}, \cite{M1}, on the classical probabilistic representation 
of QM; see also the article of  D'Ariano \cite{DM} on the information characterization of the quantum probabilistic structure.

Finally, we emphasize that 
our construction of the classical probability
space for the EPR-Bohm-Bell experiment cannot be used to support the hidden variable approach to the quantum phenomena. The classical 
random parameter $\omega$ involved in our considerations cannot be identified with the hidden variable $\lambda$ which is used 
the Bell-type considerations. Randomness of $\omega$ is not reduced to randomness of the state preparation; it also includes 
randomness of selections of experimental settings. The model is so tricky that these randomnesses cannot be separated from each other.
Therefore we are not able to determine (even probabilistically) the values of incompatible physical observables on the basis of 
``hidden variables'' associated with prepared systems.
 
Nevertheless, this work made smaller the gap between classical and quantum probabilities by interpreting the latter as classical 
conditional probabilities.  

\section{Experiments on compound systems combining a few experimental settings}
\label{EXP}

There is a source producing compound systems; we consider the simplest case of systems consisting of two subsystems, $S=(S_1, S_2).$ 
There are given two measurement devices labeled as the $A$-device and the $B$-device; the first one performs measurements on  
$S_1$ and the second on  $S_2.$ It can (but need not) be assumed that the devices
are spatially separated for sufficiently large distance.  Each device can measure a few observables, $A_i, i=1,...,n, 
B_j, j=1,...,m.$ 
Again for simplicity we restrict the number of observables per device to two, $n=m=2,$  
and consider observables taking values $\pm 1.$ (A bit later the range of values will be 
extended to include the event of nondetection.) At the fixed moment of time each device can measure
only one observable. Selections of observables to be measured are orchestrated by random generators denoted by $a$ and $b.$ In our case each takes 
two values $a, b=1,2$ determining devices' settings corresponding to measurements of observables labeled by these indexes.

Measurement of each  observable is performed with the aid of a pair of detectors, $D^A_i(-), D^A_i(+)$ and  $
D^B_i(-), D^B_i(+), i=1,2.$ At ``each side'' the corresponding random generator controls the flow of (sub)systems in the following way.
To be concrete consider the $A$-device. If the random generator $a$ takes the value $a=1,$ then the subsystem $S_1$ in the pair $S$ arrives 
to one of the detectors    $D^A_1(-), D^A_1(+)$ and produces a click. (It is assumed that all 
detectors have $100\%$ efficiency.) For this value of $a,$ there is no way for $S_1$ to arrive to one of the detectors $D^A_2(-), D^A_2(+).$ 
Moreover, there is neither random background, so no one of the detectors $D^A_2(-), D^A_2(+)$ can click.\footnote{  
As a possible realization, we can consider the following experimental framework. The random generator $a$ is coupled to a 
block which splits the channel going from the source of photons in the $A$-direction into two channels, which are also labeled by $i=1,2.$  
Each of the channels is coupled to its own polarization beam splitter (PBS) 
which has the fixed orientation given by the angle $\theta_i$
and the PBS in the $i$th channel is coupled 
to its pair of the detectors   $D^A_i(-),$ polarization down, and $D^A_i(+),$ polarization up.  Thus at each side there are two 
PBSs (corresponding to the fixed orientations) and totally 4 detectors. The complete two-side experimental scheme is based on 4 PBSs and 8 detectors.} 
If $D^A_1(-)$ clicks, we set $A_1=-1,$ and  if $D^A_1(+)$ clicks, we set $A_1=+1.$  
In this context,  $D^A_2(\pm)$ detectors do not click. We label this situation for the $A_2$-observable by setting 
$A_2=0;$ thus nondetection is labeled by zero. This is a trivial, but important point: {\it in our model the range of values 
of observables is extended by adding the value (zero) corresponding to nondetection.}  

Typically in quantum experiments the number of detectors per device is restricted to just one pair, e.g.,  the same pair of detectors is used 
to measure both $A_1$ and $A_2.$ Our model also covers this case by identification the detectors for $i=1,2.$ However, even in such a case 
it is useful to preserve the labeling by the experimental setting. In principle, change of experimental setting can modify functioning of detectors.

This is the most general description of experiments on compound systems combing selections of a few experimental settings. This scheme 
can be applied both to classical and quantum systems. In quantum optics settings are given by angles determining orientations of PBSs,
$\theta_1, \theta_2$ for the $A$-device and $\theta_1^\prime, \theta_2^\prime$ for the $B$-device. We can also consider an experiment
(classical or quantum) in which one of observables is position and another momentum.    

The main message of this section: if a few settings of measurement devices are combined in one experiment, then the rule of their 
combination has to be present in the corresponding probabilistic model; in our model this rule is given by the pair of random generators.

\section{Classical probability model describing combination of a few experimental settings}

Let $(\Omega, {\cal F}, p)$ be Kolmogorov probability space. Consider four random variables $A_i\equiv A_i(\omega)$ 
and $B_i\equiv B_i(\omega), i=1,2,$ which yield the values $\{-1,0, +1\}.$ (They describe observables.)
We also assume that on the same probability space there are defined other two random variables $a\equiv a(\omega)$ and
$b\equiv b(\omega);$ they yield the values $i=1,2.$  (They describe random generators.)  

These random variables are connected by the following conditions $(i=1,2):$
\begin{equation}
\label{L1}
{\bf C1}: p(A_i=0 \vert a\not=i)=1, p(B_i=0 \vert b\not=i)=1;
\end{equation}
\begin{equation}
\label{L2}
{\bf C2}:  p(A_i=0 \vert a=i)=0, p(B_i=0 \vert b=i)=0.
\end{equation}
Thus $A_i$ and $B_i$ can take nonzero values only under conditions $a=i$ and 
$b=i,$ respectively;  e.g.,   
\begin{equation}
\label{L3}
p(A_i =-1 \vert a=i) + p(A_i =+1\vert a=i) =1 -  p(A_i=0 \vert a=i)=1;
\end{equation} 
\begin{equation}
\label{L4}
p(A_i =-1 \vert a\not=i) + p(A_i =+1\vert a\not=i) =1 -  p(A_i=0 \vert a \not=i)=0.
\end{equation} 

These conditions have different physical meanings. By ${\bf C1}$ there is no random background. For example, for the $A$-device,
detectors in a pair coupled to the $i$-setting cannot click if systems are sent to the pair corresponding to another setting.  By 
${\bf C2}$ all detectors have $100\%$ efficiency.  Thus to model less efficient detectors, one has to proceed without ${\bf C2}.$  
(Our model can be easily modified to include both the presence of random background and usage of detectors with the efficiency less than
$100\%.$ However, we want to present the essentials of our approach for embedding quantum probabilities into the 
classical probability model; we do not want to shadow the essentials by ``technicalities".)

By using these conditions we derive (with the aid of the formula of total probability and more generally additivity of probability) 
some properties of unconditional probabilities. 

First of all we have $(i=1,2,):$
\begin{equation}
\label{L5}
p(A_i =0)=  p(a=i)p(A_i =0 \vert  a=i) + p(a\not=i) p(A_i =0 \vert  a\not=i) = p(a\not=i) .
\end{equation} 
Thus the probability that the detectors $D^A_i(\pm)$ do not click equals to the probability of non-selection of the
$i$-setting. In the same way we obtain
\begin{equation}
\label{L57}
p(B_i =0)=p(b\not=i).
\end{equation} 
By using (\ref{L4}) we find that, for $\epsilon, \epsilon^\prime= \pm 1,$    
\begin{equation}
\label{L5t}
p(A_i =\epsilon)=  p(a=i)p(A_i =\epsilon \vert  a=i)
\end{equation} 
or another way around 
\begin{equation}
\label{L5a}
p(A_i =\epsilon \vert  a=i)= \frac{1}{ p(a=i)} p(A_i =\epsilon).
\end{equation} 

As the next step we find the joint probability distribution for the pairs of the random variables $(A_i, B_j).$ Set
$\epsilon, \epsilon^\prime= \pm 1.$ Then 
\begin{equation}
\label{L6}
p(A_i =\epsilon, B_j =\epsilon^\prime)= \sum_{k,m=1,2} p(A_i =\epsilon, B_j =\epsilon^\prime, a=k, b=m).  
 \end{equation} 
Now suppose that, e.g., $k\not=i.$ Then we obtain $$
p(A_i =\epsilon, B_j =\epsilon^\prime, a=k, b=m) \leq 
p(A_i =\epsilon, a=k)= p(a\not=i) p(A_i =\epsilon\vert a\not= i)=0.$$ 
Thus only the terms with $k=i$ and $m=j$ 
give nontrivial contributions into the sum. We obtain the following formula:
\begin{equation}
\label{L7}
p(A_i =\epsilon, B_j =\epsilon^\prime)=  p(A_i =\epsilon, B_j =\epsilon^\prime \vert a=i, b=j)  p(a=i, b=j)  
 \end{equation} 
 or 
\begin{equation}
\label{L8}
p(A_i =\epsilon, B_j =\epsilon^\prime \vert a=i, b=j)=  \frac{1}{p(a=i, b=j)} p(A_i =\epsilon, B_j =\epsilon^\prime).  
\end{equation} 
Now we find joint probabilities of non-detection:
\begin{equation}
\label{L9}
p(A_i =0, B_j =0) =\sum_{k,m=1,2} p(A_i =0, B_j =0, a=k, b=m)   
 \end{equation} 
 $$
 =  p(A_i =0, B_j =0, a\not=i, b\not=j).
 $$
Hence,
\begin{equation}
\label{L10}
p(A_i =0, B_j =0) = p(A_i =0, B_j =0\vert  a\not=i, b\not=j) p(a\not=i, b\not=j).
\end{equation} 
We can also rewrite (\ref{L9}) in the following way
\begin{equation}
\label{L11}
 p(a\not=i, b\not=j\vert A_i =0, B_j =0)= 1.   
 \end{equation} 
No detection for the pair of observables $(A_i, B_j)$ implies that the pair $(i,j)$ of setting was not selected.  
In the same way we derive that (for $\epsilon, \epsilon^\prime=\pm 1$)
\begin{equation}
\label{L12}
p(A_i =\epsilon, B_j =0) = p(A_i =\epsilon, B_j =0\vert  a =i, b\not=j) p(a=i, b\not=j), 
\end{equation}
\begin{equation}
\label{L12g}
p(A_i =0, B_j =\epsilon^\prime) = p(A_i =0, B_j =\epsilon^\prime \vert  a \not=i, b=j) p(a\not=i, b=j). 
\end{equation} 

\subsection{Quantum probabilities as conditional probabilities}
\label{COND}

Now this is the time to explain why we  are interested in coupling between ``absolute probabilities'' and conditional 
probabilities. ``Absolute probabilities'' contain the contribution of randomness involved in selection of the experimental settings.
This randomness is not present in conditional probabilities. Therefore the conditional probabilities 
match with the probabilities which are obtained by experimenters
for the fixed experimental settings. If we forget (at least for a moment) about non-detection probabilities, then the basic
relation of this paper is given by the formula (\ref{L8}).

In particular, consider experiments to test the Bell-type inequalities. Here the probabilities
$p(A_i =\epsilon, B_j =\epsilon^\prime \vert a=i, b=j)$ correspond to probabilities obtained for the fixed 
pair of angles $(\theta_i, \theta_j^\prime).$  Theoretically these probabilities are predicted by the formalism of quantum mechanics.
The relation (\ref{L8}) implies the following interpretation of quantum probabilities: from the viewpoint of the classical probabilistic
model these are {\it conditional probabilities} with respect to selections of pairs of experimental settings. Of course, features of  conditional 
probabilities  do not coincide with features of absolute probabilities. Therefore attempts of interpretation of quantum probabilities as ``absolute
probabilities** induce
problems and even paradoxes. We shall consider this problem closer in section \ref{Bell}.
    
\section{Counterfactuals}

As in any classical probabilistic model, in our model there are well defined not only joint probability distributions 
for the pairs of random variables $(A_i, B_j),$ but even for the pairs $(A_i, A_j)$ and $(B_i, B_j).$ In quantum mechanics 
the corresponding observables are incompatible and cannot be measured jointly. It seems that there is a kind of contradiction.
However, this is not the case. In our model of experiment (section \ref{EXP}), e.g., observables $A_1$ and  $A_2$ cannot be 
measured jointly, although the corresponding pair of random variables is well defined, $\omega \to (A_1(\omega), A_2(\omega),$
for $\omega \in \Omega.$ It is easy to see that, for $\epsilon_1, \epsilon_2= \pm 1,$ 
\begin{equation}
\label{L5h7}
p(A_1=\epsilon_1, A_2=\epsilon_2) = 0.
\end{equation}
In other words the measure of the set 
$$
O_A(\epsilon_1 \epsilon_2)=\{\omega \in \Omega:  A_1(\omega)=\epsilon_1, A_2(\omega)
=\epsilon_2\}
$$ 
equals to zero. One can say that in our model of experiment {\it counterfactuals exist, but the probability to meet them 
equals to zero.}

To prove (\ref{L5h7}), we represent its left-hand side as
$$
p(A_1=\epsilon_1, A_2=\epsilon_2)= p(A_1=\epsilon_1, A_2=\epsilon_2, a=1) + p(A_1=\epsilon_1, A_2=\epsilon_2, a=2)
$$
$$
\leq p(A_2=\epsilon_2, a=1) + p(A_1=\epsilon_1,  a=2)
$$
$$
= p(A_2=\epsilon_2 \vert a=1) p(a=1) + p(A_1=\epsilon_1 \vert a=2) p(a=2)=0, 
$$
where at the last step the equality (\ref{L4}) was applied.

Of course, the same can be said about the probability distribution of quadruples. For $x_1, x_2, y_1, y_2= 0, \pm 1,$
if $ x_1x_2\not=0$ or $y_1y_2\not=0,$ then
\begin{equation}
\label{L5h76}
p(A_1=x_1, A_2=x_2, B_1=y_1, B_2=y_2) = 0.    
\end{equation}
Thus the measure of of the set $$O_A(x_1x_2y_1 y_2)=\{\omega \in \Omega:  A_1(\omega)=x_1, A_2(\omega)
=x_2, B_1(\omega)=y_1, B_2(\omega)=y_2 \},$$ where  $ x_1x_2\not=0$ or $y_1y_2\not=0,$  equals to zero.

\section{Classical probabilistic description of (non)locality}
\label{LOC}

We remark that in the Bell framework the notion of (non)locality was not formulated in the  space-time picture 
which is  standard for relativity theory. And in this paper we do not plan to discuss this problem, 
see, e.g., \cite{book2}, \cite{B1}--\cite{B3}) for the corresponding formulation and discussion. The original Bell discussion on 
(non)locality was given in terms of probabilistic independence and in this paper we shall proceed in the same way.
In our model Bell-type locality is formalized with the aid of the following conditions on observables (represented by 
the random variables $A_i, B_j, i,j=1,2)$ and on random generators determining experimental settings (and represented by 
the random variables $a,b).$ 

\medskip

${\bf LO}$ {\it The observables are indexed by just one index encoding to selections of the corresponding experimental setting, i.e.,
$A_i, B_j$ and not} $A_{ij}, B_{ij}.$
 
\medskip

${\bf LIG}$ {\it The random generators determining selections of the experimental settings are independent random variables:}
\begin{equation}
\label{L5h}
p(a=i, b=j) = p(a=i) p(b=j)
\end{equation}
or
\begin{equation}
\label{L5h1}
p(a=i\vert b=j)= p(a=i), 
p(b=j\vert a=i) =p(b=j).    
\end{equation} 
 
 However, it is not enough to assume that selections of the settings are done independently. One also has to be sure that 
observables ``at one side'' are independt from selections of  the settings ``at another side''. This condition jointly with 
${\bf LG}$  leads us to condition:

\medskip

${\bf LIOG}$ {\it The random vectors $(A_i,a), i=1,2,$ are  independent from  the random generator $b$  
and  the random vectors $(B_i,b), i=1,2,$ are  independent from  the random generator $a.$ }

\medskip

We remark that  ${\bf LIOG}$ trivially implies ${\bf LIG}.$ Thus we now proceed under conditions 
${\bf LO}, {\bf LIOG}$ (of course, in combination with ${\bf C1}, {\bf C2}$). In the framework of the Bell-type
tests we consider these conditions as classical probabilistic formalization of locality ( $\rm{mod}$ remark at the beginning of this 
section); violation of ${\bf LO}$ or ${\bf LIOG}$ leads to (probabilistic) nonlocality. 

\medskip

We now analyze consequences of ${\bf LIOG};$ for $x=0, \pm 1, m, k=1,2,$ we have: 
$
p(A_i=x, a=k) = p(A_i=x, a=k\vert b=m).
$
Thus 
$
p(A_i=x) = \sum_{k=1,2} p(A_i=x, a=k)= \sum_{k=1,2}p(A_i=x, a=k\vert b=m).
$ 
Hence,  
\begin{equation}
\label{L5h}
p(A_i =x \vert b=m)= p(A_i =x),
\end{equation} 
and, moreover,  $p(A_i =x \vert a=k, b=m)= p(A_i=x,  a=k, b=m)/p(a=k, b=m)=
p(A_i=x,  a=k)/p(a=k).$ Thus 
\begin{equation}
\label{L5h}
p(A_i =x \vert a=k, b=m) = p(A_i =x \vert a=k).
\end{equation} 
Thus, for measurement of a single observable, conditioning by the ``two-side selection'' of experimental settings 
is equivalent to the conditioning by the  `one-side selection.'' 

We also have:

$$
p(A_i=0, B_j=0)= p(A_i=0, B_j=0, a\not=i, b\not=j)
$$
$$
= p(A_i=0, B_j=0, a\not=i, b\not=j) 
$$
$$
+ p(A_i=0, B_j=1, a\not=i, b\not=j) +p(A_i=0, B_j=-1, a\not=i, b\not=j)
$$
$$
=p(A_i=0,  a\not=i, b\not=j)= p(A_i=0,  a\not=i) p(b\not=j)
$$
$$
=p(A_i=0\vert a\not=i) p(a\not=j) p(b\not=j).
$$
By using ${\bf C1}$ we obtain 
\begin{equation}
\label{DL1}
p(A_i=0, B_j=0)= p(a\not=j) p(b\not=j)= p(a\not=j, b\not=j).
\end{equation}
Thus the probability of the event ``no detection for the pair of observables $(A_i, B_j)$'' is equal 
to the probability of the event ``the pair of random generators did not take the values $(i, j)$''.
From (\ref{L5}) and (\ref{L57}) we find that events of nondetection in the A-device and the B-device 
are independent: 
\begin{equation}
\label{DL1}
p(A_i=0, B_j=0)= p(A_i=0) p(B_j=0).
\end{equation}
This is one of signs of ``locality'' of the probabilistic model.  In fact, even detection in one of devices is independent 
from nondetection at another, e.g., for $\epsilon = \pm 1,$
$$
p(A_i=\epsilon, B_j=0)= p(A_i=\epsilon, B_j=0, a=i, b\not=j)
$$
$$
= p(A_i=\epsilon, B_j=0, a =i, b\not=j) 
$$
$$
+ p(A_i=\epsilon, B_j=1, a=i, b\not=j) +p(A_i=\epsilon, B_j=-1, a=i, b\not=j)
$$
$$
=p(A_i=\epsilon,  a=i, b\not=j)= p(A_i=\epsilon,  a=i) p(b\not=j)
$$
$$
=p(A_i=\epsilon \vert a=i) p(a=i) p(b\not=j).
$$
Finally, we use (\ref{L5}) and (\ref{L5a}):
\begin{equation}
\label{DL2}
p(A_i=\epsilon, B_j=0)= p(A_i=\epsilon) p(B_j=0).
\end{equation}
The same is valid for the $B$-detection:
\begin{equation}
\label{DL2}
p(A_i=0, B_j=\epsilon^\prime)= p(A_i=0) p(B_j=\epsilon^\prime).
\end{equation}
In terms of Bell's argument these are also the signs of ``locality'' of the model. 

\section{Classical probabilistic description of no-signaling}

In the Kolmogorov model the probabilities for the values of a single
random variable can be reconstructed from the probabilities for the  values
of pairs of random variables. For our model, this implies that, for $j=1,2,$ 
\begin{equation}
\label{L5h77}
p(A_i =x )= \sum_y p(A_i =x, B_j=y),
\end{equation} 
where $x,y=0, \pm 1.$ This is an evident feature of a probability measure,
we call it the condition of {\it marginal consistency.} 

In quantum mechanics  the no-signaling condition 
is realization of the condition of marginal consistency 
in context of the Bell-type tests, see the work of  Kofler and Brukner  \cite{Kofler1} for detailed analysis of the problem
in the rigorous probabilistic framework.
 As was pointed out in section \ref{COND},
in our classical probabilistic model ``quantum probabilities'' appear as 
conditional probabilities (conditioning with respect to fixed 
experimental settings). In general the condition of marginal consistency 
for ``absolute probabilities'', see (\ref{L5h77}), does not coincide 
with this condition for conditional probabilities. The latter can be violated. 
So, in general ``signaling'' is possible (of course, here we discuss, 
as is common 
in the Bell framework, (no)signaling in purely probabilistic terms, see remark at the
beginning of section \ref{LOC}). However, we shall show that under the conditions 
of (probabilistic) locality, ${\bf LO}$ and ${\bf LIOG},$ the ``absolute marginal 
consistency'' (\ref{L5h77}) implies the conditional marginal consistency:
\begin{equation}
\label{L5h5}
p(A_i =\epsilon \vert a=i )= \sum_{\epsilon^\prime} p(A_i =\epsilon, B_j=\epsilon^\prime \vert a=i, b=j).
\end{equation} 
We prove this statement. The equality (\ref{L8}) implies
$$
\Sigma =\sum_{\epsilon^\prime} p(A_i =\epsilon, B_j=\epsilon^\prime \vert a=i, b=j)=
\frac{1}{p(a=i, b=j)} \sum_{\epsilon^\prime} p(A_i =\epsilon, B_j=\epsilon^\prime).
$$
For ``absolute probabilities'' we have
$$
 \sum_{\epsilon^\prime}  p(A_i =\epsilon, B_j=\epsilon^\prime) = p(A_i =\epsilon)  - p(A_i =\epsilon, B_j=0).
$$
Independence of detection in the $A$-device from nondetection in the $B$-device\footnote{The terminology 
might be misleading. To escape this problem, we recall that in our model detectors have 100\% efficiency; nondetection 
means just that this experimental setting was not selected.}, see (\ref{DL2}), implies: 
$$
\Sigma = \frac{1}{p(a=i, b=j)} (p(A_i =\epsilon)  - p(A_i =\epsilon, B_j=0))=
\frac{p(A_i =\epsilon) (1- p(B_j=0))}{p(a=i, b=j)}.
$$
Finally, we use (\ref{L57}), independence of random generators, and (\ref{L5a}):
$$
\Sigma = \frac{p(A_i =\epsilon) (1- p(b\not=j))}{p(a=i, b=j)} 
= \frac{p(A_i =\epsilon)p(b= j))}{p(a=i) p(b=j)} = p(A_i =\epsilon \vert a=i).
$$ 

We remark that the condition of marginal consistency is well known in mathematical psychology under the name 
of marginal selectivity, e.g.,  Dzhafarov et al.  \cite{DZ}, \cite{DZ1}; see Asano et al. \cite{LG} recent studies. 
    
\section{Classical probabilistic viewpoint on violation of the CHSH-inequality}
\label{Bell}

In this section we shall use the special symbol for the conditional probabilities corresponding 
to quantum probabilities, we set 
$$
q(A_i= \epsilon)\equiv  p(A_i= \epsilon\vert a=i), \; q(B_j= \epsilon^\prime)\equiv  p(B_j= \epsilon^\prime\vert b=j),
$$ 
$$
q(A_i= \epsilon, B_j= \epsilon^\prime)
\equiv  p(A_i= \epsilon\vert a=i, b=j).
$$ 

On one hand, we know that for classical probabilities the CHSH-inequality  holds; on the other hand,
we know that for quantum probabilities it can be violated. 
Consider classical correlations $(i,j=1,2):$
$$
C_{ij}= EA_i B_j= p(A_i=+1, B_j=+1) +  p(A_i=-1, B_j=-1)
$$
$$
 -  p(A_i=+1, B_j=-1) -  p(A_i=-1, B_j=+1).    
$$
They are simply connected with quantum correlations:
$$
Q_{ij}= E[A_i B_j \vert a=i, b=j]  
= \frac{C_{ij}}{p(a=i, b=j)} \; \mbox{or} \; C_{ij}=p(a=i, b=j) Q_{ij}.      
$$
The CHSH-combination of correlations is given by 
the quantity:
$$
S = C_{11}  + C_{12}  + C_{21} - C_{22}
$$
It satisfies the CHSH-inequality (which is simple theorem in the Kolmogorov probability theory):
\begin{equation}
\label{L5h28}
\vert S \vert \leq 2.
\end{equation} 
We now write this inequality in term od quantum probabilities (classical conditional probabilities):
\begin{equation}
\label{L5h28}
\vert Q_{11}p(a=1, b=1) + Q_{12} p(a=1, b=2) + Q_{21} p(a=2, b=1) - Q_{22} p(a=2, b=2)  \vert \leq 2.
\end{equation} 
Such weighted CHSH-inequality cannot be violated even by quantum correlations -- in contrast to the standard
(unweighted) CHSH-inequality which is typically used  for quantum correlations: 
\begin{equation}
\label{L5h28ttt}
\vert Q_{11} + Q_{12}  + Q_{21} - Q_{22}  \vert \leq 2,
\end{equation} 
but, as we have seen, its use is unjustified.

For example,  let $p(a=i, b=j)=1/4.$  Then (\ref{L5h28}) takes the form:
\begin{equation}
\label{L5h28y77}
\vert Q_{11} + Q_{12}  + Q_{21} - Q_{22}   \vert \leq 8.
\end{equation} 
But the number in the right-hand side,  $\rm{rhs}=8,$ is too large. Since each
quantum correlation $Q_{ij}$ is less than 1 (because even each conditional probability is probability, i.e., 
it is normalized by 1, and each of random variables $A_i, B_j$ takes the values in $[-1,+1]$), one can expect 
at least  $\rm{rhs}=4.$

The main point is that the CHSH inequality holds for an arbitrary Kolmogorov space and an arbitrary quadrupole of random 
variables taking values  in $[-1,+1].$ However, our model is based on special quadrupole of random 
variables.  We can use this specialty. It can  easily shown  
that for the ``absolute Kolmogorov correlations $C_{ij}$ a stronger inequality holds true:
\begin{equation}
\label{L5h28bb}
\vert S \vert \leq 1.
\end{equation} 
For $p(a=i, b=j)=1/4,$ it implies  that 
\begin{equation}
\label{L5h28yhh}
\vert Q_{11} + Q_{12}  + Q_{21} - Q_{22}   \vert \leq 4.
\end{equation}

\section{Conclusions}

We demonstrated that ``quantum probabilities and correlations'' can be peacefully embedded in the classical Kolmogorov model:
they have to be interpreted as {\it conditional probabilities}, where conditioning is with respect to fixed experimental
settings. This approach leads to the weighted CHSH-inequality (\ref{L5h28yhh}) (for symmetric and independent random generators
determining settings of measurement devices) and not to the standard CHSH-inequality (\ref{L5h28ttt}). In the classical probabilistic 
framework the violation of the latter can be expected. Of course, our scheme based on the complete account of randomness can be applied
to any experiment in which a few (in general incompatible) experimental settings are involved. Therefore one can expect violation 
of  (\ref{L5h28ttt}) for conditional probabilities and the corresponding conditional correlations
 arising in various experiments, i.e., not only the experiments of the 
EPR-Bohm-Bell type.

Therefore ``quantumness'' is characterized not by  a violation of the inequality (\ref{L5h28ttt}), i.e., not by the experimentally confirmed 
fact that $\vert Q_{11} + Q_{12}  + Q_{21} - Q_{22}  \vert $ can be larger than 2, but by {\it the Tsirelson bound} $2\sqrt{2}.$
It cannot be explained in the classical Kolmogorov framework.

\medskip

We also emphasize that our classical probabilistic model does not provide the objective representation of observables; although 
one can assign  simultaneously the values for both observables $A_1, A_22$ as well as 
for $B_1, B_2$ (as in any classical probabilistic model), the probability to obtain a nontrivial combination, $(\epsilon_1, \epsilon_2), 
\epsilon_i=\pm 1,$ equals to zero. Thus the possibility of the classical probabilistic description is not equivalent to the possibility
to treat observables as objective entities, cf. Zeilinger \cite{Z1}, \cite{Z} and Brukner and Zeilinger \cite{BR1}--\cite{BR3}. 
In general our model as well as the quantum 
model is not about ``reality as it is'', but about information which can be gained with the aid of observables, cf. again with   as well 
as works of  Chiribella,  D'Ariano, and Perinotti \cite{DM1} on the information approach to QM.

Although our model cannot be considering as supporting the ideas about hidden variables and in general it matches well 
with the information interpretation of QM, in the light of the existence of the classical probabilistic model
for quantum probabilistic data violating the CHSH-inequality the problem of  {\it irreducible quantum 
randomness} (from von Neumann to Brukner and Zeilinger)   has to be reanalyzed once again. Since the quantum probabilities 
and correlations can be represented as the classical entities (conditional with respect to the selections of experimental contexts),
one has to understand how conditioning can shadow the irreducible randomness. It seems that here the key word is contextuality.
Everybody would agree that there is a link (not completely understandable) between contextuality and quantum randomness. The usage 
of the specialty of quantum randomness in some schemes of quantum cryptography and theory of quantum random generators lifts 
the problem of the coexistence of quantum randomness and classical probabilistic representation to the level of quantum technologies.

\section*{acknowledgments}

This paper was written during author's visiting professor fellowship to the Institute for Quantum Optics and Quantum Information 
of Austrian Academy of Science (April-June, 2014); the main result of this paper was presented in the course of lectures on the inter-relation
between classical and quantum randomness given for the graduate students of this institute. I would like to thank Anton Zeilinger for hospitality 
and critical discussions about the objective representation of quantum observables and the possibility to apply the mathematical formalism of 
quantum mechanics  outside of quantum physics.

\end{document}